\documentclass[10pt,twoside]{article}
\usepackage{fancyhdr}
\usepackage{titlesec}
\usepackage{cite}
\usepackage{ifthen}
\usepackage{graphicx}

\titleformat{\section}{\centering\large\bfseries}{\S\arabic{section}}{1em}{}
\newboolean{first}
\setboolean{first}{true}

\begin{document}

\setlength\abovedisplayskip{2pt}
\setlength\abovedisplayshortskip{0pt}
\setlength\belowdisplayskip{2pt}
\setlength\belowdisplayshortskip{0pt}

\title{\bf \Large  An efficient dynamic ID based remote user authentication scheme using self-certified public keys for multi-server environment
\author{Dawei Zhao$^{a,b}$  \  \ Haipeng Peng$^{a,b}$   \  \ Shudong Li$^{c}$ \ \ Yixian Yang$^{a,b}$ \\ \small \it $^{a}$Information Security Center, Beijing University of Posts and
Telecommunications, \\
\small \it Beijing 100876, China. \\ \small \it $^{b}$National
Engineering Laboratory for Disaster Backup and
Recovery, \\
\small \it Beijing University of Posts and Telecommunications,
Beijing 100876, China. \\ \small \it $^{c}$ School of Mathematics,
Shandong Institute of Business and Technology, \\
\small \it Shandong Yantai, 264005 China.}\date{}} \maketitle

\footnote{E-mail address: dwzhao@ymail.com (Dawei Zhao);
penghaipeng@bupt.edu.cn (Haipeng Peng).}

\begin{center}
\begin{minipage}{135mm}
{\bf \small Abstract}.\hskip 2mm {\small Recently, Li et al.
analyzed Lee et al.'s multi-server authentication scheme and
proposed a novel smart card and dynamic ID based remote user
authentication scheme for multi-server environments. They claimed
that their scheme can resist several kinds of attacks. However,
through careful analysis, we find that Li et al.'s scheme is
vulnerable to stolen smart card and offline dictionary attack,
replay attack, impersonation attack and server spoofing attack. By
analyzing other similar schemes, we find that the certain type of
dynamic ID based multi-server authentication scheme in which only
hash functions are used and no registration center participates in
the authentication and session key agreement phase is hard to
provide perfect efficient and secure authentication. To compensate
for these shortcomings, we improve the recently proposed Liao et
al.'s multi-server authentication scheme which is based on pairing
and self-certified public keys, and propose a novel dynamic ID based
remote user authentication scheme for multi-server environments.
Liao et al.'s scheme is found vulnerable to offline dictionary
attack and denial of service attack, and cannot provide user's
anonymity and local password verification. However, our proposed
scheme overcomes the shortcomings of Liao et al.'s scheme. Security
and performance analyses show the proposed scheme is secure against
various attacks and has many excellent features.}
\end{minipage}\end{center}
\begin{center}
\begin{minipage}{135mm}
{\bf \small Keyword}.\hskip 2mm {\small Authentication,
Multi-server, Pairing-based, Hash function, Self-certified public
keys.}
\end{minipage}
\end{center}

\section{Introduction}

With the rapid development of network technologies, more and more
people begin using the network to acquire various services such as
on-line financial, on-line medical, on-line shopping, on-line bill
payment, on-line documentation and data exchange, etc. And the
architecture of server providing services to be accessed over the
network often consists of many different servers around the world
instead of just one. While enjoying the comfort and convenience of
the internet, people are facing with the emerging challenges from
the network security.

Identity authentication is the key security issue of various types
of on-line applications and service systems. Before an user
accessing the services provided by a service provider server, mutual
identity authentication between the user and the server is needed to
prevent the unauthorized personnel from accessing services provided
by the server and avoid the illegal system cheating the user by
masquerading as legal server. In the single server environment,
password based authentication scheme [1] and its enhanced version
which additionally uses smart cards [2-9] are widely used to provide
mutual authentication between the users and servers. However, the
conventional password based authentication methods are not suitable
for the multi-servers environment since each user does not only need
to log into different remote servers repetitively but also need to
remember many various sets of identities and passwords if he/she
wants to access these service providing servers. In order to resolve
this problem, in 2000, based on the difficulty of factorization and
hash function, Lee and Chang [10] proposed a user identification and
key distribution scheme which agrees with the multi-server
environment. Since then, authentication schemes for the multi-server
environment have been widely investigated and designed by many
researchers [11-28].

Based on the used of the basic cryptographic algorithms, the
existing multi-server authentication schemes can be divided into two
types, namely the hash based authentication schemes and the
public-key based authentication schemes. At the same time, among
these existing multi-server authentication schemes, some of them
need the registration center (RC) to participate in the
authentication and session key agreement phase, while others don't.
Therefore, according to the participation or not of the RC in the
authentication and session key agreement phase, we divide the
multi-server authentication schemes into RC dependented
authentication schemes and non-RC dependented authentication
schemes.

In this paper, we analyze a novel multi-server authentication
scheme, Li et al.'s scheme [20] which is only based on hash function
and a non-RC dependented authentication scheme. We find that this
scheme is vulnerable to stolen smart card and offline dictionary
attack, replay attack, impersonation attack and server spoofing
attack. By analyzing some other similar schemes [15,17-19], we find
that the type of dynamic ID based multi-server authentication scheme
which is only using hash functions and non-RC dependented is hard to
provide perfect efficient and secure authentication. To compensate
for these shortcomings, we improve the recently proposed Liao et
al.'s multi-server authentication scheme [27] which is based on
pairing and self-certified public keys, and propose a novel dynamic
ID based remote user authentication scheme for multi-server
environments. Liao et al.'s scheme is found vulnerable to offline
dictionary attack [28] and denial of service attack, and cannot
provide user's anonymity and local password verification. However,
our proposed scheme overcomes the shortcomings of Liao et al.'s
scheme. Security and performance analyses show the proposed scheme
is secure against various attacks and has many excellent features.

\section{Related works}

A large number of authentication schemes have been proposed for the
multi-server environment. Hash function is one of the key
technologies in the construction of multi-server authentication
scheme. In 2004, Juang et al. [11] proposed an efficient
multi-server password authenticated key agreement scheme based on a
hash function and symmetric key cryptosystem. In 2009, Hsiang and
Shih [12] proposed a dynamic ID based remote user authentication
scheme for multi-server environment in which only hash function is
used. However, Sood et al. [13] found that Hsiang and Shih's scheme
is susceptible to replay attack, impersonation attack and stolen
smart card attack. Moreover, the password change phase of Hsiang and
Shih's scheme is incorrect. Then Sood et al. presented a novel
dynamic identity based authentication protocol for multi-server
architecture to resolve the security flaws of Hsiang and Shih's
scheme [13]. After that, Li et al. [14] pointed out that Sood et
al.'s protocol is still vulnerable to leak-of-verifier attack,
stolen smart card attack and impersonation attack. At the same time,
Li et al. [14] proposed another dynamic identity based
authentication protocol for multi-server architecture. However, the
above mentioned scheme are all RC dependented multi-server
authentication scheme. In 2009, Liao and Wang [15] proposed a
dynamic ID based multi-server authentication scheme which is based
on hash function and non-RC dependented. But, Liao and Wang's scheme
is vulnerable to insider's attack, masquerade attack, server
spoofing attack, registration center spoofing attack and is not
reparable [16]. After that, Shao et al. [17] and Lee et al. [18,19]
proposed some similar types of multi-server authentication schemes.
In 2012, Li et al.[20] pointed out that Lee et al.'s scheme [18]
cannot withstand forgery attack, server spoofing attack and cannot
provide proper authentication, and then proposed a novel dynamic ID
based multi-server authentication schemes which is only using hash
function and non-RC dependented. However, with careful analysis, we
find that Li et al.'s scheme [20] is still vulnerable to stolen
smart card and offline dictionary attack, replay attack,
impersonation attack and server spoofing attack. We also analyzed
Shao et al.'s scheme [17] and Lee et al.'s scheme [19], they are all
vulnerable to stolen smart card and offline dictionary attack,
replay attack, impersonation attack and server spoofing attack. In
general, it is difficult to construct a secure dynamic ID based and
non-RC dependented multi-server authentication scheme if only hash
functions are used.

Public-key cryptograph is another useful technique which is widely
used in the construction of multi-server authentication scheme. In
2000, Lee and Chang [21] proposed a user identification and key
distribution scheme in which the difficulty of factorization on
public key cryptography is used. In 2001, Tsaur [22] proposed a
remote user authentication scheme based on RSA cryptosystem and
Lagrange interpolating polynomial for multi-server environments.
Then Lin et al. [23] proposed a multi-server authentication protocol
based on the simple geometric properties of the Euclidean and
discrete logarithm problem concept. Since the traditionally public
key cryptographic algorithms require many expensive computations and
consume a lot of energy, Geng and Zhang [24] proposed a dynamic
ID-based user authentication and key agreement scheme for
multi-server environment using bilinear pairings. But Geng and
Zhang's scheme cannot withstand user spoofing attack [25]. After
that, Tseng et al. [26] proposed an efficient pairing-based user
authentication scheme with smart cards. However, in 2013, Liao and
Hsiao [27] pointed out that Tseng et al.'s scheme is vulnerable to
insider attack, offline dictionary attack and malicious server
attack, and cannot provide proper mutual authentication and session
key agreement. At the same time, Liao and Hsiao proposed a novel
non-RC dependented multi-server remote user authentication scheme
using self-certified public keys for mobile clients [27]. Recently,
Chou et al. [28] found Liao and Hsiao's scheme cannot withstand
password guessing attack. Furthermore, with careful analysis, we
find that Liao and Hsiao's scheme is still vulnerable to denial of
service attack, and cannot provide user's anonymity and local
password verification. In this paper, based on the Liao and Hsiao's
scheme, we propose a secure dynamic ID based and non-RC dependented
multi-server authentication scheme using the pairing and
self-certified public keys.

\section{Review and cryptanalysis of Li et
al.'s authentication scheme}
\subsection{Review of Li et al.'s scheme} \label{}

Li et al.'s contains three participants, the user $U_i$, the server
$S_j$, and the registration center $RC$. $RC$ chooses the master
secret key $x$ and a secret number $y$ to compute $h(x\|y)$ and
$h(SID_j\|h(y))$, and then shares them with $S_j$ via a secure
channel. $SID_j$ is the identity of server $S_j$. There are four
phases in the scheme: registration phase, login phase, verification
phase, and password change phase.

\subsubsection{Registration phase} \label{}

When the remote user authentication scheme starts, the user $U_i$
and the registration center $RC$ need to perform the following steps
to finish the registration phase:

(1) $U_i$ freely chooses his/her identity $ID_i$, the password
$PW_i$, and computes $A_i = h(b\oplus PW_i)$, where $b$ is a random
number generated by $U_i$. Then $U_i$ sends $ID_i$ and $A_i$ to the
registration center $RC$ for registration through a secure channel.

(2) $RC$ computes $B_i = h(ID_i\| x)$, $C_i = h(ID_i\|h(y)\|A_i)$,
$D_i = h(B_i \| h(x \| y))$ and $E_i = B_i\oplus h(x \| y)$. $RC$
stores $\{C_i, D_i, E_i, h(\cdot), h(y)\}$ on the user's smart card
and sends it to user $U_i$ via a secure channel.

(3) $U_i$ keys $b$ into the smart card, and finally the smart card
contains $\{C_i, D_i, E_i, b, h(\cdot), h(y)\}$.

\subsubsection{Login phase} \label{}
 Whenever $U_i$ wants to login $S_j$, he/she must perform the
following steps to generate a login request message:

(1) $U_i$ inserts his/her smart card into the card reader and inputs
$ID_i$ and $PW_i$. Then the smart card computes $A_i = h(b\oplus
PW_i)$, $C_i^* = h(ID_i\|h(y)\|A_i)$, and checks whether the
computed $C_i^*$ is equal to $C_i$. If they are equal, $U_i$
proceeds the following steps. Otherwise the smart card aborts the
session.

(2) The smart card generates a random number $N_i$ and computes
$P_{ij} = E_i\oplus h(h(SID_j\|h(y))\|N_i)$, $CID_i = A_i\oplus
h(D_i\|SID_j\|N_i)$, $M_1 = h(P_{ij}\|CID_i\|D_i\|N_i)$ and $M_2 =
h(SID_j\|h(y))\oplus N_i$.

(3) $U_i$ submits $\{P_{ij}, CID_i,M_1,M_2\}$ to $S_j$ as a login
request message.

\subsubsection{Verification phase} \label{}

Wher $S_j$ receiving the login message $\{P_{ij}, CID_i,M_1,M_2\}$,
$S_j$ and $U_i$ perform the following steps to finish the mutual
authentication and session key agreement.

(1) $S_j$ computes $N_i = M_2\oplus h(SID_j\|h(y))$, $E_i = P_{ij}
\oplus h(h(SID_j\|h(y))\|N_i)$, $B_i = E_i\oplus h(x\|y)$, $D_i =
h(B_i\|h(x\|y))$ and $A_i = CID_i\oplus h(D_i\|SID_j\|N_i)$ by using
$\{P_{ij}, CID_i,M_1,M_2\}$, $h(SID_j\|h(y))$ and $h(x\|y)$.

(2) $S_j$ computes $h(P_{ij}\|CID_i\|D_i\|N_i)$ and checks whether
it is equal to $M_1$. If they are not equal, $S_j$ rejects the login
request and terminates this session. Otherwise, $S_j$ accepts the
login request message. Then $S_j$ generates a random number $N_j$
and computes $M_3 = h(D_i\|A_i\|N_j\|SID_j)$, $M_4 = A_i\oplus
N_i\oplus N_j$. Finally, $S_j$ sends the message $\{M_3,M_4\}$ to
$U_i$.

(3) After receiving the response message $\{M_3,M_4\}$ sent from
$S_j$, $U_i$ computes $N_j = A_i\oplus N_i\oplus M_4$,
$M_3^*=h(D_i\|A_i\|N_j \|SID_j)$ and checks $M_3^*$ with the
received message $M_3$. If they are not equal, $U_i$ rejects these
messages and terminates this session. Otherwise, $U_i$ successfully
authenticates $S_j$. Then, the user $U_i$ computes the mutual
authentication message $M_5 = h(D_i\|A_i\|N_i\|SID_j)$ and sends
$\{M_5\}$ to the server $S_j$.

(4) Upon receiving the message $\{M_5\}$ from $U_i$, $S_j$ computes
$h(D_i\|A_i\|N_i\|SID_j)$ and checks it with the received message
$\{M_5\}$. If they are equal, $S_j$ successfully authenticates $U_i$
and the mutual authentication is completed. After the mutual
authentication phase, the user $U_i$ and the server $S_j$ compute
$SK = h(D_i\|A_i\|N_i\|N_j\|SID_j)$, which is taken as their session
key for future secure communication.

\subsubsection{Password change phase} \label{}

This phase is invoked whenever $U_i$ wants to change his password
$PW_i$ to a new password $PW^{new}_i $. There is no need for a
secure channel for password change, and it can be finished without
communicating with the registration center $RC$.

(1) $U_i$ inserts his/her smart card into the card reader and inputs
$ID_i$ and $PW_i$.

(2) The smart card computes $A_i = h(b\oplus PW_i)$, $C^*_i =
h(ID_i\|h(y)\|A_i)$, and checks whether the computed $C^*_i$ is
equal to $C_i$. If they are not equal, the smart card rejects the
password change request. Otherwise, the user $U_i$ inputs a new
password $PW^{new}_i$ and a new random number $b^{new}$.

(3) The smart card computes $A^{new}_i = h(b^{new}\oplus PW^{new}_i
)$ and $C^{new}_i = h(ID_i\|h(y)\|A^{new}_i)$.

(4) Finally, the smart card replaces $C_i$ and $b$ with $C^{new}_i$
and $b^{new}$ to finish the password change phase.

\subsection{Cryptanalysis of Li et al.'s scheme} \label{}

Li et al. claimed that their scheme can resist many types of attacks
and satisfy all the essential requirements for multi-server
architecture authentication. However, if we assume that $A$ is an
adversary who has broken a user $U_m$ and a server $S_n$, or a
combination of a malicious user $U_m$ and a dishonest server $S_n$.
Then $A$ could get the secret number $h(x\|y)$ and $h(y)$, and can
perform the stolen smart card and offline dictionary attack, replay
attack, impersonation attack and server spoofing attack to Li et
al.'s scheme. The concrete cryptanalysis of the Li et al.'s scheme
is shown as follows.

\subsubsection{Stolen smart card  and offline dictionary attack} \label{}

If a user $U_i$'s smart card is stolen by an adversary $A$, $A$ can
extract the information $\{C_i, D_i, E_i, b,$ $h(\cdot), h(y)\}$
from the memory of the stolen smart card. Furthermore, in case $A$
intercepts a valid login request message $\{P_{ij}, CID_i,M_1,M_2\}$
sent from user $U_i$ to server $S_j$ in the public communication
channel, $A$ can compute $N_i = h(SID_j\|h(y))\oplus M_2$, $E_i =
P_{ij} \oplus h(h(SID_j\|h(y))\|N_i)$, $B_i = E_i\oplus h(x\|y)$,
$D_i = h(B_i\|h(x\|y))$ and $A_i = CID_i\oplus h(D_i\|SID_j\|N_i)$
by using $h(y)$ and $h(x\|y)$. Then  $A$ can launch offline
dictionary attack on $C_i = h(ID_i\|h(y)\|A_i)$ to know the identity
$ID_i$ of the user $U_i$ because $A$ knows the values of $A_i$
corresponding to the user $U_i$. Besides $A$ can launch offline
dictionary attack on $A_i = h(b\oplus PW_i)$ to know the password
$PW_i$ of $U_i$ because $A$ knows the value of $b$ from the stolen
smart card of the user $U_i$. Now $A$ possesses the valid smart card
of user $U_i$, knows the identity $ID_i$, password $PW_i$
corresponding to the user $U_i$ and hence can login on to any
service server.

\subsubsection{Replay attack} \label{}

The replay attack is replaying the same message of the receiver or
the sender again. If adversary $A$ has intercepted a valid login
request message $\{P_{ij}, CID_i,M_1,M_2\}$ sent from user $U_i$ to
server $S_j$ in the public communication channel. Then $A$ can
compute $N_i = h(SID_j\|h(y))\oplus M_2$, $E_i = P_{ij} \oplus
h(h(SID_j\|h(y))\|N_i)$, $B_i = E_i\oplus h(x\|y)$, $D_i =
h(B_i\|h(x\|y))$ and $A_i = CID_i\oplus h(D_i\|SID_j\|N_i)$ by using
$h(y)$ and $h(x\|y)$. Then adversary $A$ can replay this login
request message $\{P_{ij}, CID_i,M_1,M_2\}$ to $S_j$ by masquerading
as the user $U_i$ at some time latter. After verification of the
login request message, $S_j$ computes $M_3 =
h(D_i\|A_i\|N_j\|SID_j)$ and $M_4 = A_i\oplus N_i\oplus N_j$, and
sends the message $\{M_3,M_4\}$ to $A$ who is masquerading as the
user $U_i$. The adversary $A$ can verify the received value of
$\{M_3,M_4\}$ and compute $M_5' = h(D_i\|A_i\|N_i\|SID_j)$ since he
knows the values of $N_i, E_i, B_i, D_i$ and $A_i$. Then $A$ sends
$\{M'_5\}$ to the server $S_j$. The $S_j$ computes
$h(D_i\|A_i\|N_i\|SID_j)$ and checks it with the received message
$\{M_5'\}$. This equivalency authenticates the legitimacy of the
user $U_i$, the service provider server $S_j$ and the login request
is accepted. Finally after mutual authentication, adversary $A$
masquerading as the user $U_i$ and the server $S_j$ agree on the
common session key as $SK = h(D_i\|A_i\|N_i\|N_j\|SID_j)$.
Therefore, the adversary $A$ can masquerade as user $U_i$ to login
on to server $S_j$ by replaying the same login request message which
had been sent from $U_i$ to $S_j$.

\subsubsection{Impersonation attack} \label{}

In this subsection, we show that the adversary $A$ who possesses
$h(y)$ and $h(x \| y)$ can masquerade as any user $U_i$ to login any
server $S_j$ as follows.

Adversary $A$ chooses two random numbers $a_i$ and $b_i$, and
computes $A_i=h(a_i)$ and $B_i=h(b_i)$. Then $A$ can compute $D_i =
h(B_i \| h(x \| y))$, $E_i = B_i\oplus h(x \| y)$, $P_{ij} =
E_i\oplus h(h(SID_j\|h(y))\|N_i)$, $CID_i = A_i\oplus
h(D_i\|SID_j\|N_i)$, $M_1 = h(P_{ij}\|CID_i\|D_i\|N_i)$ and $M_2 =
h(SID_j\|h(y))\oplus N_i$ by using $h(y)$ and $h(x \| y)$. Now $A$
sends the login request message $\{P_{ij}, CID_i,M_1,M_2\}$ by
masquerading as the user $U_i$ to server $S_j$. After receiving the
login request message, $S_j$ computes $N_i = h(SID_j\|h(y))\oplus
M_2$, $E_i = P_{ij} \oplus h(h(SID_j\|h(y))\|N_i)$, $B_i = E_i\oplus
h(x\|y)$, $D_i = h(B_i\|h(x\|y))$ and $A_i = CID_i\oplus
h(D_i\|SID_j\|N_i)$ by using $\{P_{ij}, CID_i,M_1,M_2\}$, $h(x\|y)$
and $h(SID_j\|h(y))$. Then $S_j$ computes $M_3 =
h(D_i\|A_i\|N_j\|SID_j)$ and $M_4 = A_i\oplus N_i\oplus N_j$, and
sends the message $\{M_3,M_4\}$ to $A$ who is masquerading as the
user $U_i$. Then adversary $A$ computes $N_j = A_i\oplus N_i\oplus
M_4$ and verifies $M_3$ by computing $h(D_i\|A_i\|N_j\|SID_j)$. Then
$A$ computes $M_5 = h(D_i\|A_i\|N_i\|SID_j)$ and sends $\{M_5\}$
back to the server $S_j$. The $S_j$ computes
$h(D_i\|A_i\|N_i\|SID_j)$ and checks it with the received message
$\{M_5\}$. This equivalency authenticates the legitimacy of the user
$U_i$, the service provider server $S_j$ and the login request is
accepted. Finally after mutual authentication, adversary $A$
masquerading as the user $U_i$ and the server $S_j$ agree on the
common session key as $SK = h(D_i\|A_i\|N_i\|N_j\|SID_j)$.

\subsubsection{Server spoofing attack} \label{}

In this subsection, we show that the adversary $A$ who possesses
$h(y)$ and $h(x \| y)$ can masquerade as the server $S_j$ to spoof
user $U_i$, if $A$ has intercepted a valid login request message
$\{P_{ij}, CID_i,M_1,M_2\}$ sent from user $U_i$ to server $S_j$ in
the public communication channel.

After intercepting a valid login request message $\{P_{ij},
CID_i,M_1,M_2\}$ sent from user $U_i$ to server $S_j$ in the public
communication channel, $A$ can compute $N_i = h(SID_j\|h(y))\oplus
M_2$, $E_i = P_{ij} \oplus h(h(SID_j\|h(y))\|N_i)$, $B_i = E_i\oplus
h(x\|y)$, $D_i = h(B_i\|h(x\|y))$ and $A_i = CID_i\oplus
h(D_i\|SID_j\|N_i)$ corresponding to $U_i$. Then $A$ can choose a
random number $N'_j$, and compute $M_3 = h(D_i\|A_i\|N'_j\|SID_j)$
and $M_4 = A_i\oplus N_i\oplus N'_j$. $A$ then sends the message
$\{M_3,M_4\}$ by masquerading as server $S_j$ to the user $U_i$.
After receiving the message $\{M_3,M_4\}$, $U_i$ computes $N'_j =
A_i\oplus N_i\oplus M_4$ and verifies $M_3$ by computing
$h(D_i\|A_i\|N'_j\|SID_j)$. Then $U_i$ computes $M_5 =
h(D_i\|A_i\|N_i\|SID_j)$ and sends it to the $S_j$ who is
masquerading as the adversary $A$. Then $A$ computes
$h(D_i\|A_i\|N_i\|SID_j)$ and checks it with the received message
$\{M_5\}$. Finally after mutual authentication, adversary $A$
masquerading as the server $S_j$ and the user $U_i$ agree on the
common session key as $SK = h(D_i\|A_i\|N_i\|N'_j\|SID_j)$.

\subsection{Discussion} \label{}

Except the Li et al.'s scheme, we also analyzed other four dynamic
ID based authentication schemes for multi-server environment
[15,17-19]. These schemes are all based on hash functions and non-RC
dependented. We found that such type of multi-server remote user
authentication scheme are almost vulnerable to stolen smart card and
offline dictionary attacks, impersonation attack and server spoofing
attack etc. The cryptanalysis methods of these schemes are similar
to that of Li et al.'s scheme shown in section 3.2. We think that
under the assumptions that no registration center participates in
the authentication and session key agreement phase, the dynamic ID
and hash function based user authentication schemes for multi-server
environment is hard to provide perfect efficient and secure
authentication. Fortunately, there is another technique, public-key
cryptograph which is widely used in the construction of
authentication scheme. Therefore, in order to construct a secure,
low power consumption and non-RC dependented authentication scheme,
we adopt the elliptic curve cryptographic technology of public-key
techniques, and propose a novel dynamic ID based and non-RC
dependented remote user authentication scheme using pairing and
self-certified public keys for multi-server environment.

\section{Preliminaries} \label{}

Before presenting our scheme, we introduce the concepts of bilinear
pairings, self-certified public keys, as well as some related
mathematical assumptions.

\subsection{Bilinear pairings} \label{}

Let $G_1$ be an additive cyclic group with a large prime order $q$
and $G_2$ be a multiplicative cyclic group with the same order $q$.
Particularly, $G_1$ is a subgroup of the group of points on an
elliptic curve over a finite field $E(F_p)$ and $G_2$ is a subgroup
of the multiplicative group over a finite field. $P$ is a generator
of $G_1$.

A bilinear pairing is a map $e : G_1\times G_1\rightarrow G_2$ and
satisfies the following properties:

(1) Bilinear: $e(aP, bQ)=e(P, Q)^{ab}$ for all $P, Q\in G_1$ and $a,
b\in  Z^*_q$.

(2) Non-degenerate: There exists $P, Q\in G_1$ such that $e(P,
Q)\neq 1$.

(3) Computability: There is an efficient algorithm to compute $e(P,
Q)$ for all $P, Q\in G_1$.

\subsection{self-certified public keys} \label{}

In [27], Liao et al. first proposes a key distribution based on
self-certified public keys (SCPKs) [29,30] among the service
servers. By using the SCPK, a user's public key can be computed
directly from the signature of the third trust party (TTP) on the
user's identity instead of verifying the public key using an
explicit signature on a user's public key. The SCPK scheme is
described as follows.

(1) Initialization: The third trust party (TTP) first generates all
the needed parameters of the scheme. TTP chooses a non-singular high
elliptic curve $E(F_p)$ defined over a finite field, which is used
with a based point generator $P$ of prime order $q$. Then TTP freely
chooses his/her secret key $s_T$ and computes his/her public key
$pub_T = s_T\cdot P$. The related parameters and $pub_T$ are
publicly and authentically available.

(2) Private key generation: An user $A$ chooses a random number
$k_A$, computes $K_A=k_A\cdot P$ and sends his/her identity $ID_A$
and $K_A$ to the TTP. TTP chooses a random number $r_A$, computes
$W_A = K_A + r_A\cdot P$ and $\bar{s}_A=h(ID_A\parallel W_A)+r_A$,
and sends $W_A$ and $\bar{s}_A$ to user $A$. Then $A$ obtains
his/her secret key by calculating $s_A=\bar{s}_A+k_A$.

(3) Public key extraction: Anyone can calculate $A$'s public key
$pub_A=h(ID_A\parallel W_A)pub_T+ W_A$ when he/she receives $W_A$.

\subsection{Related mathematical assumptions} \label{}

To prove the security of our proposed protocol, we present some
important mathematical problems and assumptions for bilinear
pairings defined on elliptic curves. The related concrete
description can be found in [31,32].

(1) Computational discrete logarithm (CDL) problem: Given $R =
x\cdot P$, where $P, R \in  G_1$. It is easy to calculate $R$ given
$x$ and $P$, but it is hard to determine $x$ given $P$ and $R$.

(2) Elliptic curve factorization (ECF) problem: Given two points $P$
and $R= x\cdot P +y\cdot P$ for $x, y\in Z^*_q$ , it is hard to find
$x\cdot P$ and $y\cdot P$.

(3) Computational Diffie-Hellman (CDH) problem: Given $P, xP, yP \in
G_1$, it is hard to compute $xyP\in  G_1$.

\section{The proposed scheme}
\label{}

In this section, by improving the recently proposed Liao et al.'s
multi-server authentication scheme [27] which is found vulnerable to
offline dictionary attack and denial of service attack [28], and
cannot provide user's anonymity and local password verification, we
propose a novel dynamic ID based remote user authentication scheme
for multi-server environment using pairing and self-certified public
keys. Our scheme contains three participants: the user $U_i$, the
service provider server $S_j$, and the registration center $RC$. The
legitimate user $U_i$ can easily login on to the service provider
server using his smart card, identity and password. There are six
phases in the proposed scheme: system initialization phase, the user
registration phase, the server registration phase, the login phase,
the authentication and session key agreement phase, and the password
change phase. The notations used in our proposed scheme are
summarized in Table 1.

\begin{table}\centering{
\caption{Notations used in the proposed scheme.}
\label{tab:1}       
\begin{tabular}{lllll}
\hline\noalign{\smallskip}

$e$ & & A bilinear map, $e: G_1\times G_1 \longrightarrow G_2$.\\
$U_i$ & & The $i$th user.\\
$ID_i$ & & The identity of the user $U_i$.\\
$S_j$ & & The $j$th service provider server.\\
$SID_j$ & & The identity of the service provider server $S_j$.\\
$RC$ & & The registration center. \\
$s_{RC}$ & & The master secret key of the registration center $RC$
in $Z_q^*$.\\
$pub_{RC}$ & & The public key of $RC$, $pub_{RC} = s_{RC}\cdot P$.\\
$P$ & & A generator of group $G_1$.\\
$H( )$ & & A map-to-point function, $H : {0, 1}^*\longrightarrow
G_1$.\\
$h( )$ & & A one way hash function, $h : {0, 1}^*\longrightarrow {0,
1}^k$, where $k$ is the\\ & &  output length. $h( )$ allows the
concatenation of some integer\\ & & values and points on an elliptic
curve.\\
$\oplus$ & & A simple XOR operation in $G_1$. If $P_1, P_2 \in G_1$,
$P_1$ and $P_2$ are\\ & &  points on an elliptic curve over a finite
field, the operation\\ & &  $P_1\oplus P_2$ means that it performs
the XOR operations of the\\ & &  x-coordinates and y-coordinates of
$P_1$ and
$P_2$, respectively.\\
$\parallel$ & & The concatenation operation.\\
\noalign{\smallskip}\hline
\end{tabular}}
\end{table}

\subsection{System initialization phase} \label{}

In the proposed scheme, registration center $RC$ is assumed a third
trust party. In the system initialization phase, $RC$ generates all
the needed parameters of the scheme.

(1) $RC$ selects a cyclic additive group $G_1$ of prime order $q$, a
cyclic multiplicative group $G_2$ of the same order $q$, a generator
$P$ of $G_1$, and a bilinear map $e:G_1\times G_1\longrightarrow
G_2$.

(2) $RC$ freely chooses a number $s_{RC}\in Z_q^*$ keeping as the
system private key and computes $pub_{RC} = s_{RC}\cdot P$ as the
system public key.

(3) $RC$ selects two cryptographic hash functions $H(\cdot)$ and
$h(\cdot)$.

Finally, all the related parameters $\{e, G_1, G_2, q, P, Pub_{RC},
H(\cdot), h(\cdot)\}$ are publicly and authentically available.

\subsection{User registration phase} \label{}

When the user $U_i$ wants to access the services, he/she has to
submit his/her some related information to the registration center
$RC$ for registration. The steps of the user registration phase are
as follows:

(1) The user $U_i$ freely chooses his/her identity $ID_i$ and
password $pw_i$, and chooses a random number $b_i$. Then $U_i$
computes $HPW_i=h(ID_i\parallel pw_i\parallel b_i)\cdot P$, and
submits $ID_i$ and $HPW_i$ to $RC$ for registration via a secure
channel.

(2) When receiving the message $ID_i$ and $HPW_i$, $RC$ computes
$QID_i=H(ID_i)$, $CID_i=s_{RC}\cdot QID_i$, $Reg_{ID_i}=CID_i\oplus
s_{RC}\cdot HPW_i$ and $H_i=h(QID_i
\parallel CID_i)$. Then $RC$ stores the message $\{Reg_{ID_i}, H_i\}$
in $U_i$'s smart card and submits the smart card to $U_i$ through a
secure channel.

(3) After receiving the smart card, $U_i$ enters $b_i$ into the
smart card. Finally, the smart card contains parameters
$\{Reg_{ID_i}, H_i, b_i\}$.

\subsection{Server registration phase} \label{}

If a service provider server $S_j$ wants to provides services for
the users, he/she must perform the registration to the registration
center $RC$ to become a legal service provider server. The process
of server registration phase of the proposed scheme is based on SCPK
mentioned in section 4.2.

(1) $S_j$ chooses a random number $v_j$ and computes $V_j=v_j\cdot
P$. Then $S_j$ submits $SID_j$ and $V_j$ to $RC$ for registration
via a secure channel.

(2) After receiving the message $\{SID_j, V_j\}$, $RC$ chooses a
random number $w_j$, and computes $W_j=w_j\cdot P+ V_j$ and
$s'_j=(s_{RC}\cdot h(SID_j\parallel W_j)+w_j) $ mod $q$. Then $RC$
submits the message $\{W_j, s'_j\}$ to $S_j$ through a secure
channel.

(3) After receiving $\{W_j, s'_j\}$, $S_j$ computes the private key
$s_j = (s'_j + v_j)$ mod $q$, and checks the validity of the values
issued to him/her by checking the following equation:
$pub_j=s_j\cdot P = h(SID_j\parallel W_j)\cdot pub_{RC}+W_j$. At
last, $S_j$'s personal information contains $\{SID_j, pub_j, s_j,
W_j\}$

The details of user registration phase and server registration phase
are shown in Fig.1.

\begin{figure}[h]
\centering{\includegraphics[scale=0.7,trim=0 0 0 0]{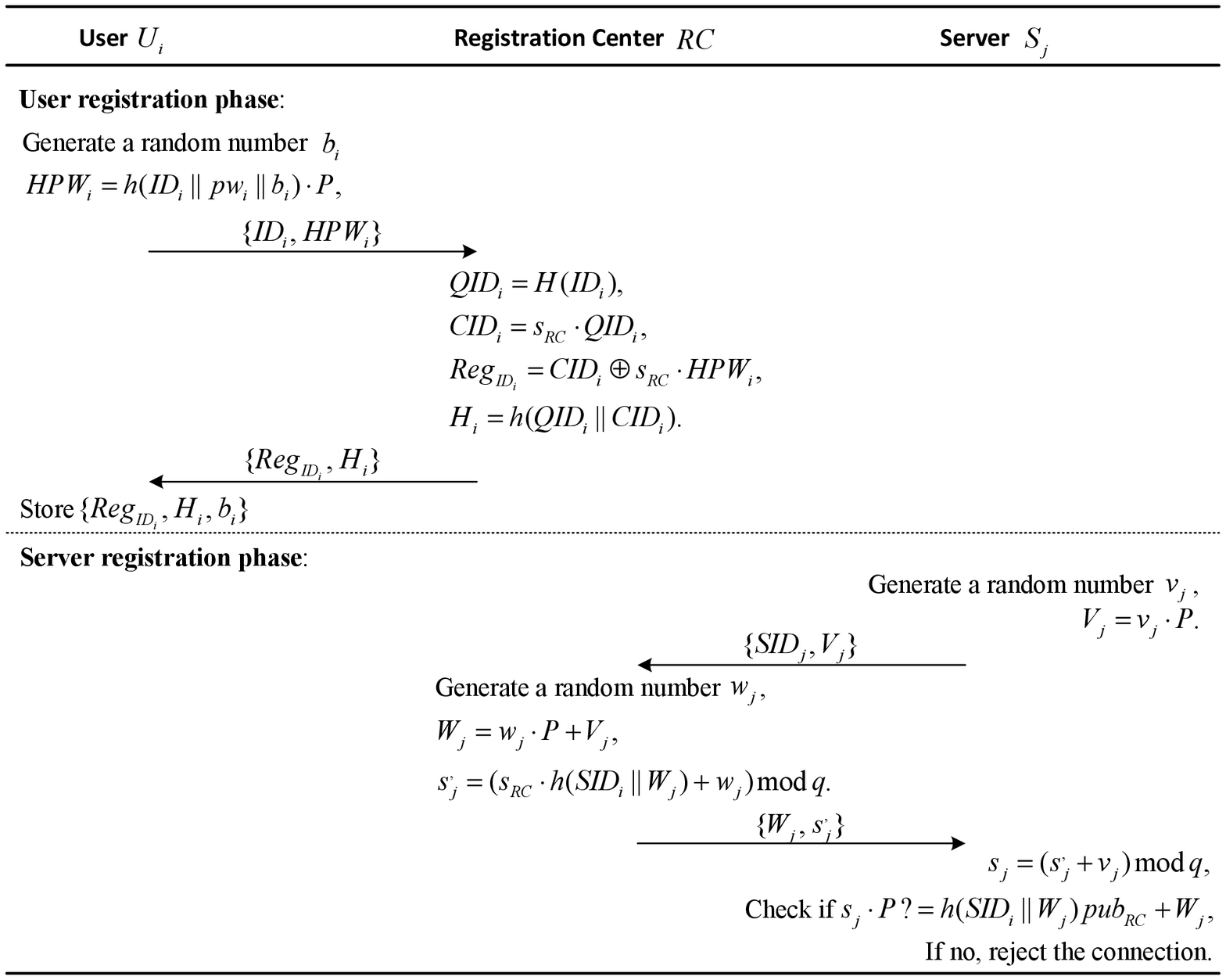}}
\caption{User and server registration phase of the proposed scheme.}
\end{figure}

\subsection{Login phase} \label{}

If user $U_i$ wants to access the services provided by server $S_j$,
$U_i$ needs to login on to $S_j$, the process of the login phase are
as following:

(1) $U_i$ inserts his/her smart card into the smart card reader, and
inputs identity $ID_i$ and password $pw_i$. Then the smart card
computes $QID_i=H(ID_i)$, $CID_i=Reg_{ID_i} \oplus h(ID_i\parallel
pw_i\parallel b_i)\cdot pub_{RC}$, $H^*_i=h(QID_i\parallel CID_i)$,
and checks whether $H^*_i=H_i$. If they are equal, it means $U_i$ is
a legal user. Otherwise the smart card aborts the session.

(2) The smart card generates two random numbers $u_i$ and $r_i$, and
computes $DID_i=u_i\cdot QID_i$ and $R_i=r_i\cdot P$. Then the smart
card sends the login request message $\{DID_i, R_i\}$ to server
$S_j$ over a public channel.

\subsection{Authentication and session key agreement phase}
\label{}

(1) After receiving the login request $\{DID_i, R_i\}$ sent from
$U_i$, $S_j$ chooses a random number $r_j$, and computes
$R_j=r_j\cdot P$, $T_{ji}=r_j\cdot R_i$, $K_{ji}=s_j\cdot R_i$ and
$Auth_{ji}=h(DID_i\parallel SID_j\parallel K_{ji}\parallel R_j)$.
Then $S_j$ sends the message $\{W_j, R_j, Auth_{ji}\}$ to $U_i$.

(2) When receiving $\{W_j, R_j, Auth_{ji}\}$, $U_i$ computes
$T_{ij}=r_i\cdot R_j$, $pub_j=h(SID_j\parallel W_j)\cdot
pub_{RC}+W_j$, $K_{ij}=r_i\cdot pub_j$ and
$Auth_{ij}=h(DID_i\parallel SID_j\parallel K_{ij}\parallel R_j)$.
Then $U_i$ checks $Auth_{ij}$ with the received $Auth_{ji}$. If they
are not equal, $U_i$ terminates this session. Otherwise, $S_j$ is
authenticated, and $U_i$ continues to compute $M_i=r_i\cdot DID_i$,
$N_i=u_i\cdot CID_i$, $d_{ij}=h(DID_i\parallel SID_j\parallel
K_{ij}\parallel M_i)$ and $B_i=(r_i+d_{ij})\cdot N_i$. Finally,
$U_i$ sends the message $\{M_i, B_i\}$ to $S_j$.

(3) After receiving the message $\{M_i, B_i\}$ sent from $U_i$,
$S_j$ computes $d_{ji}=h(DID_i\parallel SID_j\parallel
K_{ji}\parallel M_i)$ and checks whether $e(M_i+d_{ji}\cdot DID_i,
pub_{RC})=e(B_i, P)$. If they are not equal, $S_j$ terminates this
session. Otherwise, $U_i$ is authenticated.

Finally, the user $U_i$ and the server $S_j$ agree on a common
session key as $U_i: SK=h(DID_i\parallel SID_j\parallel
K_{ij}\parallel T_{ij})$, $ S_j: SK=h(DID_i\parallel SID_j\parallel
K_{ji}\parallel T_{ji})$.

The login phase and authentication and session key agreement phase
are depicted in Fig.2.

\begin{figure}[h]
\centering{\includegraphics[scale=0.7,trim=0 0 0 0]{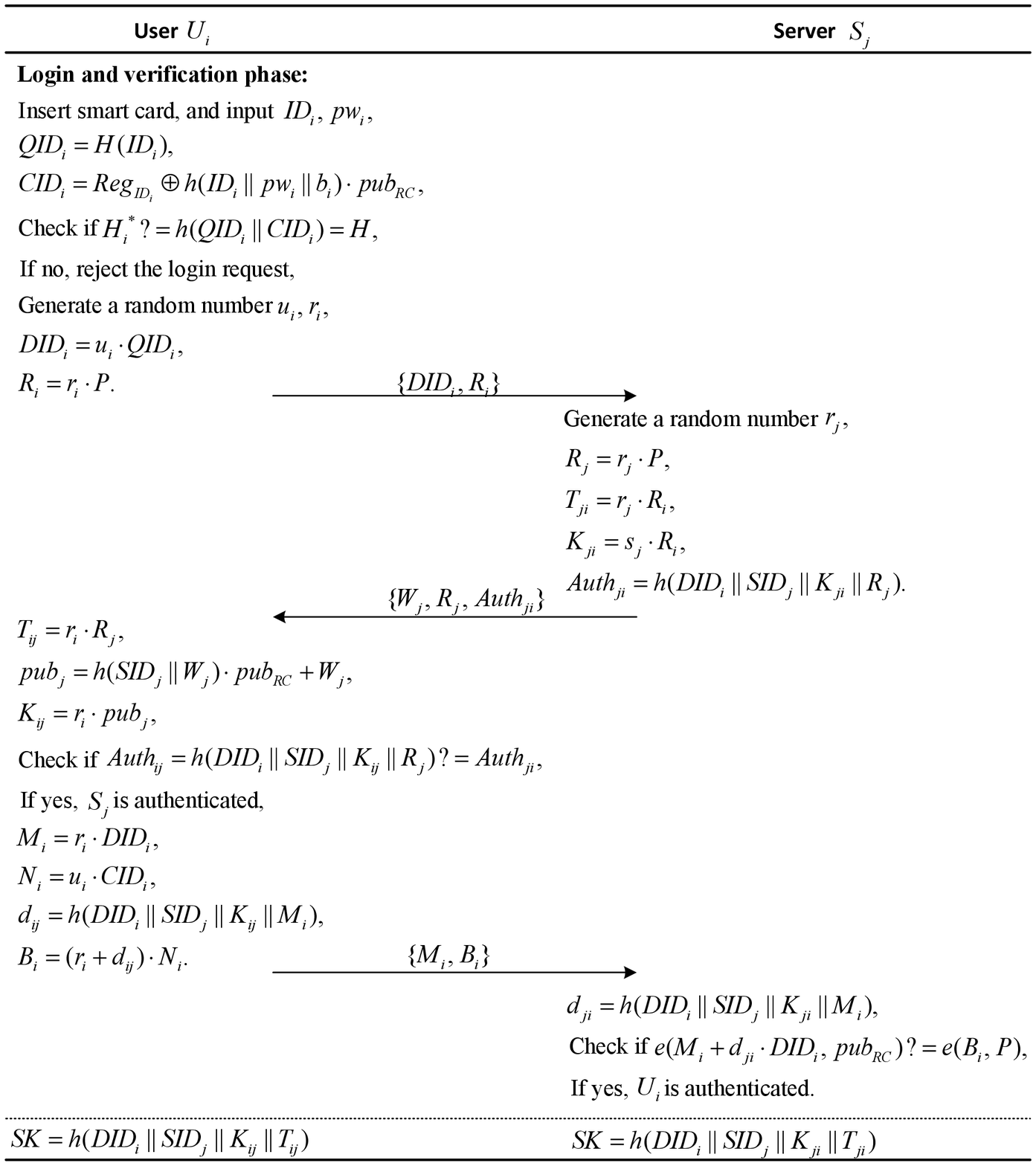}}
\caption{Login and verification phase of the proposed scheme.}
\end{figure}

\subsection{Password change phase} \label{}

The following steps show the process of the password change phase of
a user $U_i$.

(1) The user $U_i$ inserts his/her smart card into the smart card
reader, and inputs identity $ID_i$ and password $pw_i$. Then the
smart card computes $QID_i=H(ID_i)$, $CID_i=Reg_{ID_i} \oplus
h(ID_i\parallel pw_i\parallel b_i)\cdot pub_{RC}$,
$H^*_i=h(QID_i\parallel CID_i)$, and checks whether $H^*_i=H_i$. If
they are equal, it means $U_i$ is a legal user. Otherwise the smart
card aborts the session.

(2) The smart card generates a random number $z_i$, and computes
$Z_i=z_i\cdot P$ and $AID_i=CID_i\oplus z_i\cdot pub_{RC}$. Then the
smart card sends the message $\{ID_i, AID_i, Z_i\}$ to the
registration center $RC$.

(3) After receiving the message $\{ID_i, AID_i, Z_i\}$, $RC$
computes $CID_i=AID_i\oplus s_{RC}\cdot Z_i$, $QID_i=H(ID_i)$, and
checks whether $e(CID_i, P)=e(QID_i, pub_{RC})$. If they are equal,
user $U_i$ is authenticated. Then $RC$ computes
$V_1=h(CID_i\parallel s_{RC}\cdot Z_i)$ and sends $\{V_1\}$ to
$U_i$.

(4) When receiving $\{V_1\}$, user computes $h(CID_i\parallel
z_i\cdot pub_{RC})$ and checks it with the received $V_1$. If they
are equal, the registration center $RC$ is authenticated. Then $U_i$
chooses his/her new password $pw_i^{new}$ and the new random number
$b_i^{new}$, and computes $HPW_i^{new}=h(ID_i\parallel
pw_i^{new}\parallel b_i^{new})\cdot P$, $V_2=HPW_i^{new}\oplus
z_i\cdot pub_{RC}$ and $V_3=h(CID_i\parallel z_i\cdot
pub_{RC}\parallel HPW_i^{new})$. Then $U_i$ submits $\{V_2, V_3\}$
to $RC$.

(5) Upon receiving the response $\{V_2, V_3\}$, the registration
server $RC$ computes $HPW_i^{new}=V_2\oplus s_{RC}\cdot Z_i$ and
$V_3^*=h(CID_i\parallel s_{RC}\cdot Z_i\parallel HPW_i^{new})$. Then
$RC$ compares $V_3^*$ with the received $V_3$. If they are equal,
$RC$ continues to compute $Reg_{ID_i}^{new}=CID_i\oplus s_{RC}\cdot
HPW_i^{new}$, $V_4=Reg_{ID_i}^{new}\oplus s_{RC}\cdot Z_i$ and
$V_5=h(s_{RC}\cdot Z_i\parallel Reg_{ID_i}^{new})$. After that, $RC$
sends $\{V_4, V_5\}$ to $U_i$.

(6) After receiving $\{V_4, V_5\}$, $U_i$ computes
$Reg_{ID_i}^{new}=V_4\oplus z_i\cdot pub_{RC}$ and $V_5^*=h(z_i\cdot
pub_{RC}\parallel Reg_{ID_i}^{new})$. Then $U_i$ checks whether
$V_5^*=V_5$. If they are equal, user $U_i$ replaces the original
$Reg_{ID_i}$ and $b_i$ with $Reg_{ID_i}^{new}$ and $b_i^{new}$.

The details of a password change phase of the proposed scheme are
shown in Fig.3.

\begin{figure}[h]
\centering{\includegraphics[scale=0.7,trim=0 0 0 0]{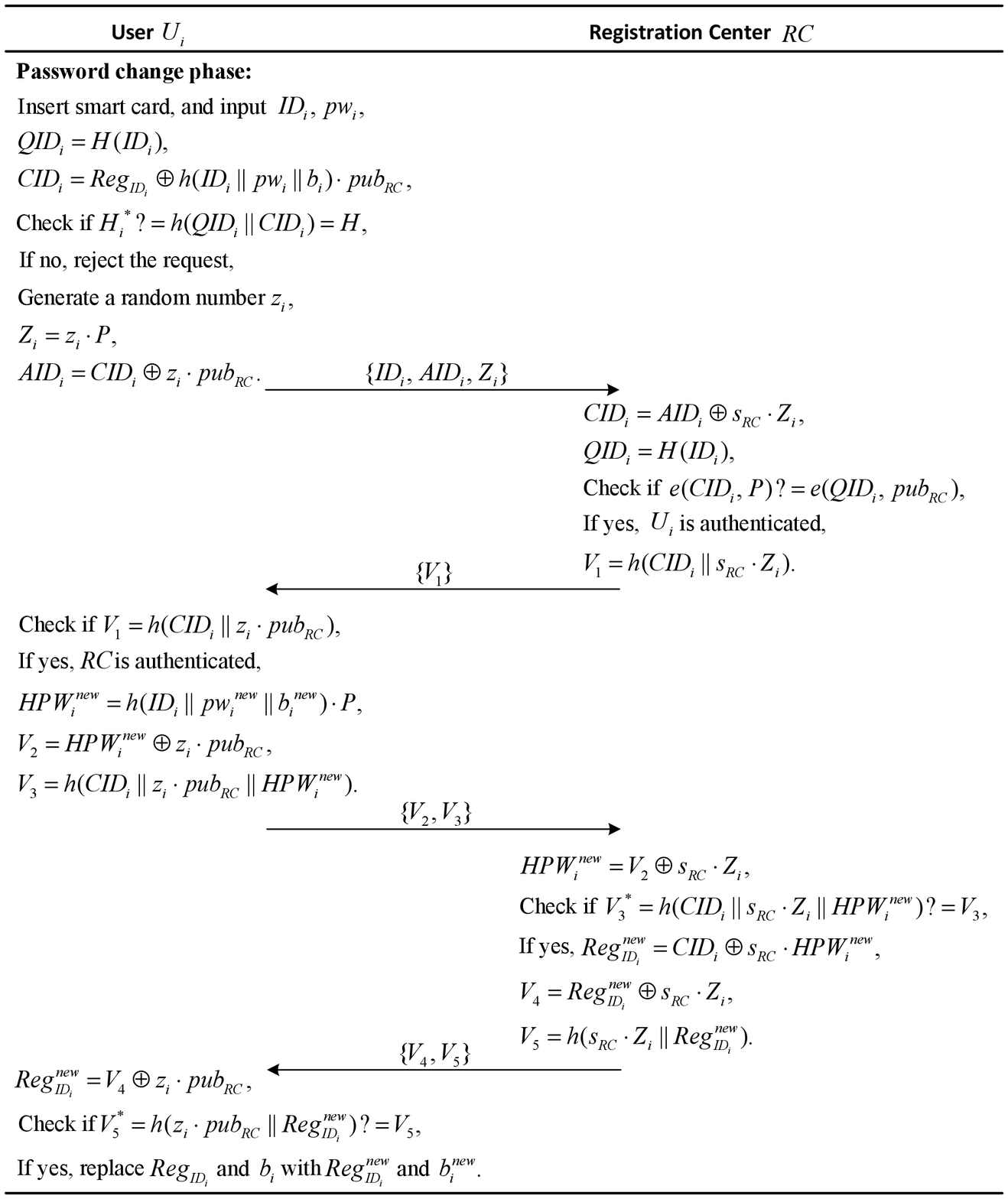}}
\caption{Password change phase of the proposed scheme.}
\end{figure}

\section{Security analysis} \label{}

\subsection{Stolen smart card and offline dictionary attacks}
\label{}

In the proposed scheme, we assume that if a smart card is stolen,
physical protection methods cannot prevent malicious attackers to
get the stored secure elements. At the same time, adversary $A$ can
access to a big dictionary of words that likely includes user's
password and intercept the communications between the user and
server.

In the proposed scheme, in case a user $U_i$'s smart card is stolen
by an adversary $A$, he can extract $\{Reg_{ID_i}, H_i\}$ from the
memory of the stolen smart card. At the same time, it is assumed
that adversary $A$ has intercepted a previous full session messages
$\{DID_i, R_i, W_j, R_j, Auth_{ji}, M_i, B_i\}$ between the user
$U_i$ and server $S_j$. However, the adversary still cannot obtain
the $U_i$'s identity $ID_i$ and password $pw_i$ except guessing
$ID_i$ and $pw_i$ at the same time. Therefore, it is impossible to
get the $U_i$'s identity $ID_i$ and password $pw_i$ from stolen
smart card and offline dictionary attack in our proposed scheme.

\subsection{Replay attack} \label{}

Replaying a message of previous session into a new session is
useless in our proposed scheme because user's smart card and the
server choose different rand numbers $r_i$ and $r_j$, and the
user'identity is different in each new session, which make all
messages dynamic and valid for that session only. If we assume that
an adversary $A$ replies an intercepted previous login request
$\{DID_i, R_i\}$ to $S_j$, after receiving the response message
$\{W_j, R_j, Auth_{ji}\}$ sent from $S_j$, $A$ cannot compute the
correct response message $\{M_i, B_i\}$ to pass the $S_j$'s
authentication since he does not know the values of $ID_i$, $pw_i$,
$u_i$ and $r_i$. Therefore, the proposed scheme is robust for the
replay attack.

\subsection{Impersonation attack} \label{}

If an adversary $A$ wants to masquerade as a legal user $U_i$ to
pass the authentication of a server $S_j$, he must have the values
of both $QID_i$ and $CID_i$. However, $QID_i$ and $CID_i$ are
protected by $U_i$'s smart card, $ID_i$ and $pw_i$ since
$QID_i=H(ID_i)$ and $CID_i=Reg_{ID_i} \oplus h(ID_i\parallel
pw_i\parallel b_i)\cdot pub_{RC}$. Therefore, unless the adversary
$A$ can obtain the $U_i$'s smart card, $ID_i$ and $pw_i$ at the same
time, the proposed scheme is secure to the impersonation attack.

\subsection{Server spoofing attack} \label{}

If an adversary $A$ wants to masquerade as a legal server $S_j$ to
cheat a user $U_i$, he must calculate a valid $Auth_{ji}$ which is
embedded with the shared secret key $K_{ji}=s_j\cdot R_i$ to pass
the authentication of $U_i$. However, adversary $A$ cannot derive
the shared secret key $K_{ji}$ without knowing the private key $s_j$
of the server $S_j$. Therefore, our scheme is secure against the
server spoofing attack.

\subsection{Insider attack} \label{}

In the proposed scheme, the registration center $RC$ cannot obtain
the $U_i$'s password $pw_i$. Since in the registration phase, $U_i$
chooses a random number $b_i$ and sends $ID_i$ and
$HPW_i=h(ID_i\parallel pw_i\parallel b_i)\cdot P$ to $RC$, $RC$ can
not derive $pw_i$ from $HPW_i$ based on CDL problem. Therefore, the
proposed scheme is robust for insider attack.

\subsection{Denial of service attack} \label{}

In denial of service attack, an adversary $A$ updates identity and
password verification information on smart card to some arbitrary
value and hence legitimate user cannot login successfully in
subsequent login request to the server. In the proposed scheme,
smart card checks the validity of user $U_i$'s identity $ID_i$ and
password $pw_i$ before password update procedure. An adversary can
insert the stolen smart card of the user $U_i$ into smart card
reader and has to guess the identity $ID_i$ and password $pw_i$
correctly corresponding to the user $U_i$. Since the smart card
computes $H^*_i=h(QID_i\parallel CID_i)$, and compares it with the
stored value of $H_i$ in its memory to verify the legitimacy of the
user $U_i$ before smart card accepts password update request. It is
not possible to guess identity $ID_i$ and password $pw_i$ correctly
at the same time in real polynomial time even after getting the
smart card of the user $U_i$. Therefore, the proposed scheme is
secure against the denial of service attack.

\subsection{Perfect forward secrecy} \label{}

Perfect forward secrecy means that even if an adversary compromises
all the passwords of the users, it still cannot compromise the
session key. In the proposed scheme, the session key
$SK=h(DID_i\parallel SID_j\parallel K_{ij}\parallel T_{ij})$ ($
SK=h(DID_i\parallel SID_j\parallel K_{ji}\parallel T_{ji})$) is
generated by three one-time random numbers $u_i$, $r_i$ and $r_j$ in
each session. These one-time random numbers are only held by the
user $U_i$ and the server $S_j$, and cannot be retrieved from $SK$
based on the security of CDH problem. Thus, even if an adversary
obtains previous session keys, it cannot compromise other session
key. Hence, the proposed scheme achieves perfect forward secrecy.

\subsection{User's anonymity} \label{}

In our proposed scheme, the user $U_i$'s login message is different
in each login phase. Among each login message, $DID_i=u_i\cdot
H(ID_i)$ is associated with a random number $u_i$ which is known by
$U_i$ only. Therefore, any adversary cannot identity the real
identity of the logon user and our scheme can provide the user's
anonymity.

\subsection{No verification table} \label{}

In our proposed scheme, it is obvious that the user, the server and
the registration center do not maintain any verification table.

\subsection{Local password verification} \label{}

In the proposed scheme, smart card checks the validity of user
$U_i$'s identity $ID_i$ and password $pw_i$ before logging into
server $S_j$. Since the adversary cannot compute the correct $CID_i$
without the knowledge of $ID_i$ and $pw_i$ to pass the verification
equation $H^*_i=H_i$, thus our scheme can avoid the unauthorized
accessing by the local password verification.

\subsection{Proper mutual authentication} \label{}

In our scheme, the user first authenticates the server. $U_i$ sends
the message $\{DID_i, R_i\}$ to the server $S_j$ to build an
connection. After receiving the response message $\{W_j, R_j,
Auth_{ji}\}$ sent from $S_j$, $U_i$ computes $T_{ij}$, $pub_j$,
$K_{ij}$, $Auth_{ij}$, and checks whether $Auth_{ij}=Auth_{ji}$. If
they are equal, $S_j$ is authenticated by $U_i$. Otherwise, $U_i$
stops to login onto this server. Since $Auth_{ji}=h(DID_i\parallel
SID_j\parallel K_{ji}\parallel R_j)$ and $K_{ji}=s_j\cdot R_i$, an
adversary $A$ cannot compute the correct $K_{ji}$ without the
knowledge of value of $s_j$. Any fabricated message $\{W'_j, R'_j,
Auth'_{ji}\}$ cannot pass the verification. Then $U_i$ computes
$M_i$, $N_i$, $d_{ij}$, $B_i$, and sends the message $\{M_i, B_i\}$
to $S_j$. After receiving the message $\{M_i, B_i\}$ sent from
$U_i$, $S_j$ computes $d_{ji}$ and checks whether $e(M_i+d_{ji}\cdot
DID_i, pub_{RC})=e(B_i, P)$. If they are not equal, $S_j$ terminates
this session. Otherwise, $U_i$ is authenticated. Since
$B_i=(r_i+d_{ij})\cdot N_i$, an adversary $A$ cannot compute the
correct $B_i$ without the knowledge of values of $u_i$ and $r_i$
etc. Any fabricated message $\{M'_i, B'_i\}$ cannot pass the
verification. Therefore, our proposed scheme can provide proper
mutual authentication.

\section{Performance comparison and functionality analysis}
\label{}

In this section, we compares the performance and functionality of
our proposed scheme with some previously schemes. To analyze the
computation cost, some notations are defined as follows.

$TG_{e}$: The time of executing a bilinear map operation, $e:
G_1\times G_1\longrightarrow G_2$.

$TG_{mul}$: The time of executing point scalar multiplication on the
group $G_1$.

$TG_H$: The time of executing a map-to-point hash function H(.).

$TG_{add}$: The time of executing point addition on the group $G_1$.

$T_h$: The time of executing a one-way hash function $h(.)$.

Since the XOR operation and the modular multiplication operation
require very few computations, it is usually negligible considering
their computation cost.

Table 2 shows the performance comparisons of our proposed scheme and
some other related protocols. We mainly focus on three computation
costs including: C1, the total time of all operations executed in
the user registration phase; C2, the total time spent by the user
during the process of login phase and verification phase; C3, the
total time spent by the server during the process of verification
phase. As shown in Table 2, Tseng et al.'s scheme are more efficient
in terms of computation cost. However, Tseng et al.'s scheme is
vulnerable to stolen smart card and offline dictionary attacks,
server spoofing attack and insider attack, and cannot provide
perfect forward secrecy, user's anonymity, proper mutual
authentication and session key agreement. In our proposed scheme,
the total computation cost of the user (C2) is
9$TG_{mul}$+$TG_H$+$TG_{add}$+5$T_h$. But similar to that in Liao et
al.'s scheme, the user $U_i$ can pre-compute $R_i = r_i\cdot P$ in
the client, and then the computation cost of the user (C2) requires
8$TG_{mul}$+$TG_H$+$TG_{add}$+5$T_h$ on-line computation. It can be
found that our proposed scheme spends a little more computation cost
than Liao et al.'s scheme in C2, and the others are almost equal.
However, Liao et al.'s scheme is vulnerable to stolen smart card and
offline dictionary attacks and denial of service attack, and cannot
provide user's anonymity and local password verification.

\begin{table}\centering{
\caption{Computational cost comparison of our scheme and other
schemes.}
\label{tab:1}       
\begin{tabular}{lllllll}
\hline\noalign{\smallskip} & Proposed scheme & Liao et al.'scheme [27] & Tseng et al.'scheme [26] & \\
\hline\noalign{\smallskip}
C1& 3$TG_{mul}$+$TG_{H}$+2$T_h$ & 3$TG_{mul}$+$TG_{H}$+$T_h$ & 2$TG_{mul}$+$TG_{H}$+$T_h$\\
C2& 8$TG_{mul}$+$TG_H$+$TG_{add}$+5$T_h$ & 5$TG_{mul}$+$TG_{H}$+$TG_{add}$+5$T_h$ & 3$TG_{mul}$+2$T_h$\\
C3& 2$TG_{e}$+4$TG_{mul}$+$TG_{add}$+2$T_h$ & 2$TG_{e}$+5$TG_{mul}$+$TG_{add}$+2$T_h$ & 2$TG_{e}$+$TG_{mul}$+$TG_{H}$+$TG_{add}$+$T_h$\\
\noalign{\smallskip}\hline
\end{tabular}}
\end{table}

Table 3 lists the functionality comparisons among our proposed
scheme and other related schemes. It is obviously that our scheme
has many excellent features and is more secure than other related
schemes.

\begin{table}\centering{
\caption{Functionality comparisons among related multi-server
authentication protocols.}
\label{tab:1}       
\begin{tabular}{llllllllll}
\hline\noalign{\smallskip}
  & Proposed & Liao & Tseng & Li & Lee & Shao & Lee\\
  & scheme & et al. & et al. & et al. & et al. & et al. & et al. \\
  &  & [27] & [26] & [20] & [18] & [17] & [19] \\
\hline\noalign{\smallskip} Resist stolen smart card and & Yes & No &
No & No & No & No & No
\\ offline dictionary attacks\\
Resist replay attack &Yes& Yes & Yes & No & No & No & No \\
Resist impersonation attack &Yes& Yes &Yes & No & No & No & No \\
Resist server spoofing attack &Yes& Yes & No & No & No & No & No \\
Resist insider attack &Yes& Yes & No & Yes & Yes & No & Yes \\
Resist denial of service attack &Yes& No & Yes & Yes & Yes & Yes & No \\
Perfect forward secrecy &Yes& Yes & No & Yes & Yes & No & No \\
User's anonymity &Yes& No & No & Yes & Yes & No & Yes \\
No verification table &Yes&  Yes  & Yes & Yes & Yes & Yes & Yes \\
Local password verification &Yes& No & Yes & Yes & Yes & Yes & No \\
Proper mutual authentication &Yes& Yes & No & Yes & No & Yes & Yes \\
\noalign{\smallskip}\hline
\end{tabular}}
\end{table}

\section{Conclusion}

In this paper, we point out that Li et al.'s scheme is vulnerable to
stolen smart card and offline dictionary attack, replay attack,
impersonation attack and server spoofing attack. Furthermore, by
analyzing some other similar schemes, we find the certain type of
dynamic ID based and non-RC dependented multi-server authentication
scheme in which only hash functions are used is hard to provide
perfect efficient and secure authentication. To compensate for these
shortcomings, we improve the Liao et al.'s multi-server
authentication scheme which is based on pairing and self-certified
public keys, and propose a novel dynamic ID based and non-RC
dependented remote user authentication scheme for multi-server
environments. The security and performance analyses show the
proposed scheme is secure against various attacks and has many
excellent features.

\section{Acknowledgment}

This paper was supported by the National Natural Science Foundation
of China (Grant Nos. 61070209, 61202362, 61121061), and the Asia
Foresight Program under NSFC Grant  (Grant No. 61161140320).

\end{document}